\documentclass[runningheads]{llncs}
\usepackage[T1]{fontenc}
\usepackage{graphicx}
\usepackage{algpseudocode}
\usepackage{algorithm}
\usepackage{comment}
\usepackage[acronym]{glossaries}
\usepackage{siunitx}
\usepackage{color,soul}
\usepackage{url}
\makeglossaries
\newacronym{cv}{CV}{computer vision}
\newacronym{nid}{NID}{Network Intrusion Detection}
\newacronym{ad}{AD}{anomaly detection}
\newacronym{nids}{NIDS}{Network Intrusion Detection System}
\newacronym{ssl}{SSL}{Self-Supervised Learning}
\newacronym{kssl}{SSL}{self-supervised learning}
\newacronym[plural=CNNs,firstplural=Convolutional Neural Networks (CNNs)]{cnn}{CNN}{Convolutional Neural Network}
\newacronym[plural=MLPs,firstplural=multilayer perceptrons (MLPs)]{mlp}{MLP}{multilayer perceptron}
\newacronym{5g}{5G-NIDD}{}
\newacronym{unsw}{UNSW-NB15}{}
\newacronym{ft}{FT-Transformer}{Feature Tokenizer
Transformer}
\newacronym{f-norm}{F-normed}{feature-wise normalized} 
\newacronym{mse}{mse}{mean-squared error}

\begin{document}
\title{An Investigation into the Performance of Non-Contrastive Self-Supervised Learning Methods for Network Intrusion Detection \thanks{Hamed Fard and Gerhard Wunder were supported by the Federal Ministry of Education and Research of Germany (BMBF) in the program of "Souverän". Digital. Vernetzt.", joint projects "UltraSec: Sicherheitsarchitektur für eine UWB-basierte Anwendungsplattform" under project identification number 16KIS1682, and "6G-RIC: 6G Research and Innovation Cluster" under project identification number 16KISK025.}}
%\thanks{Supported by organization x.}
\titlerunning{Non-Contrastive SSL for NIDS}
%
%\titlerunning{Abbreviated paper title}
% If the paper title is too long for the running head, you can set
% an abbreviated paper title here
%
\author{Hamed Fard\inst{1}\orcidID{0009-0007-2365-4313} \and
Tobias Schalau\inst{1}\orcidID{0000-0001-6881-1852} \and
Gerhard Wunder\inst{1}\orcidID{0009-0001-0850-8816}}
\authorrunning{H. Fard et al.}
% First names are abbreviated in the running head.
% If there are more than two authors, 'et al.' is used.
%
\institute{Freie Universität, Berlin Germany \\
\email{\{h.habibi.fard,g.wunder\}@fu-berlin.de}} 

\maketitle              % typeset the header of the contribution
\begin{abstract}

Network intrusion detection, a well-explored cybersecurity field, has predominantly relied on supervised learning algorithms in the past two decades. However, their limitations in detecting only known anomalies prompt the exploration of alternative approaches. Motivated by the success of self-supervised learning in computer vision, there is a rising interest in adapting this paradigm for network intrusion detection. While prior research mainly delved into contrastive self-supervised methods, the efficacy of non-contrastive methods, in conjunction with encoder architectures serving as the representation learning backbone and augmentation strategies that determine what is learned, remains unclear for effective attack detection. This paper compares the performance of five non-contrastive self-supervised learning methods using three encoder architectures and six augmentation strategies. Ninety experiments are systematically conducted on two network intrusion detection datasets, \acrshort{unsw} and \acrshort{5g}. For each self-supervised model, the combination of encoder architecture and augmentation method yielding the highest average precision, recall, F1-score, and AUCROC is reported. Furthermore, by comparing the best-performing models to two unsupervised baselines, DeepSVDD, and an Autoencoder, we showcase the competitiveness of the non-contrastive methods for attack detection. Code at: \url{https://github.com/renje4z335jh4/non_contrastive_SSL_NIDS}

\keywords{Network Intrusion Detection  \and Self-Supervised Learning \and Data Augmentation.}
\end{abstract}

\section{Introduction}
\label{Introduction}
In the face of a rising tide of security threats targeting the internet and computer networks, the need for developing flexible and adaptive security approaches is of paramount importance. The swift evolution of network technologies has increased the complexity and severity of attacks \cite{CyberEdgeGroup}. In light of this dynamic landscape, the adoption of \acrfull{nids} \cite{denning1987intrusion} has become prevalent as an effective strategy to counter the expanding threat scenario. \acrshort{nids} that use supervised methods have been the subject of extensive research over the past two decades \cite{garcia2009anomaly,yang2022systematic}. 
 However, the requirement for extensive labeled data in training poses challenges due to cost and time implications. \footnote{Alternatively, semi-supervised learning methods have shown promising performance while utilizing a few labeled samples \cite{apruzzese2022sok}. However, this paper focuses on \acrlong{kssl}, predicated on the assumption of label-free training.}
 
 \acrfull{ssl} has become increasingly prominent in recent years, offering an effective remedy for the labeled data scarcity challenge across various domains. By leveraging the underlying structure and patterns in the data, \acrshort{ssl} learns meaningful representations without the need for labeled data. Subsequently, these acquired representations are useful in other downstream tasks, such as anomaly detection \cite{hojjati2022self}. Many recent \acrshort{ssl} methods, particularly those relying on a joint-embedding architecture, share a common goal: learning representations that remain invariant under various distortions (data augmentations). In other words, these methods seek to generate similar embeddings for different augmented views of the same sample \cite{weng2022investigation}. Contrastive methods define positive and negative sample pairs through data augmentation, seeking to bring the output embeddings of positive pairs into closer proximity while simultaneously pushing negative pairs further apart \cite{chen_simple_2020}. This process requires comparing each sample with many others to work effectively. However, discarding negative samples and solely minimizing the distance between positive pairs during training can lead to the learned representation collapsing into a constant solution, where all inputs map to the same output \cite{balestriero2023cookbook}. To overcome this limitation, the computer vision (CV) community has introduced a set of \acrshort{ssl} models, including BYOL \cite{grill_bootstrap_2020}, SimSiam \cite{chen_exploring_2020}, Barlow Twins \cite{zbontar2021barlow}, VICReg \cite{bardes2021vicreg}, and W-MSE \cite{ermolov2021whitening}. These models are collectively referred to as non-contrastive methods because they require no negative samples, differing primarily in how they avoid representation collapse. Nonetheless, non-contrastive \acrshort{ssl} models rely on two crucial elements: a) data augmentations, which are essential for regulating the degree of invariance beneficial for downstream tasks, and b) encoder architectures that act as the representation learning backbone in \acrshort{ssl} models. Unlike CV, where specific data augmentations or encoder architectures have been established, and different non-contrastive \acrshort{ssl} models are commonly compared to each other \cite{bardes2021vicreg}, to the best of our knowledge, the practice of comparing these models, along with determining suitable augmentation strategies or encoder architectures for non-contrastive \acrshort{ssl} in \acrshort{nid}, remains unclear. Therefore, this paper aims to investigate the interplay between augmentation, encoder, and non-contrastive \acrshort{ssl} models by proposing a two-stage pipeline. In the initial stage, an informative representation of only normal network traffic data without labels is learned, involving augmentations, encoders, and non-contrastive \acrshort{ssl} models. In the second stage, a K-means detector is introduced to distinguish between benign and attack data. For augmentation, four strategies from prior \acrshort{nid} research and two new strategies from the tabular domain, not previously applied in \acrshort{nid}, are utilized. Regarding the encoders, three different architectures serve as the representation learning backbone for the five non-contrastive models mentioned above. Ninety different combinations are systematically investigated, stemming from the permutations of augmentation techniques, encoder architectures, and non-contrastive \acrshort{ssl} models. The best results are reported, with the most suitable combination of augmentation methods and encoder architectures being determined for each non-contrastive \acrshort{ssl} model. Performance is assessed using the metrics precision, recall, F1-score, and AUCROC on two publicly available \acrshort{nid} datasets. To the best of our knowledge, we are the first to conduct a comparative analysis of non-contrastive \acrshort{ssl} models in \acrshort{nid}, specifically examining their performance under different augmentation methods and encoder architectures. 
 
 The paper is structured as follows: In Section \ref{Related Work}, we review previous studies, emphasizing similarities and distinctions between our work and prior research. Section \ref{Method} outlines the augmentation methods, encoder architectures, \acrshort{ssl} models, and the K-means detector employed in our study. Our experimental setup is detailed in Section \ref{Experiment}. The results of our study are presented in Section \ref{Results}. The paper concludes with Section \ref{Conclusion}. 

\section{Related Work}
\label{Related Work}

In this section, prior related research on non-contrastive \acrshort{ssl} is discussed with a focus on the used \acrshort{ssl} model, augmentation strategy, and encoder architecture.

Wang et al. \cite{wang_network_2021} were the first to adopt the BYOL model from \acrshort{cv} for the task of \acrshort{nid}. Their approach involves transforming network traffic samples into grayscale images, incorporating standard computer vision augmentations such as flipping, cropping, and introducing a method called \textit{Random Shuffle}, which shuffles values within each sample. However, the authors did not provide a clear demonstration of the specific benefits of this augmentation strategy for the detection task. They also argued against using \textit{Gaussian Noise} for augmentation. In addition, the paper investigates the impact of six different encoders, all rooted in the field of \acrshort{cv}, and notes that the BoTNet attains the highest performance metric. The direct application of a CV pipeline for \acrshort{nid} has also been questioned in \cite{lotfi_network_2023}, where a zero-masking strategy (also referred to as \textit{Zero Out Noise}) is proposed for augmentation instead of relying solely on CV augmentations. BYOL is also employed in android malware detection, as demonstrated in \cite{yang2022android}, utilizing a TextCNN as an encoder and incorporating two augmentation strategies: \textit{Gaussian Noise} and row- or column-wise feature masking. Recently, BYOL has been adapted for the encrypted network classification task in \cite{towhid2022encrypted}, where data augmentation operates by dividing a flow of packets into sub-flows and using one sub-flow as an augmented version of another. These sub-flows are created through an incremental sampling strategy described in \cite{towhid2022encrypted}. Apart from BYOL, VICReg is the only other non-contrastive \acrshort{ssl} model applied in the domain of \acrshort{nid} in \cite{menon2023vicra}, where the authors used the\textit{Swap Noise} augmentation strategy from the tabular domain \cite{ucar_subtab_2021}. This technique involves randomly swapping a small portion of the columns between two samples to generate noisy augmented samples for training. For the encoder, an \acrshort{mlp} is employed.

In summary, previous research exploring the performance of non-contrastive \acrshort{ssl} models for \acrshort{nid} commonly employed a single model.  They either replicated an entire \acrshort{cv} pipeline or adopted a singular augmentation strategy along with a lone encoder. In addition, these studies consistently incorporated a supervised linear classifier in their detection stage, implying that access to a labeled subset of the dataset was assumed during the finetuning stage. In contrast, this paper conducts a) a comparative analysis of the performance among different non-contrastive \acrshort{ssl} models using six distinct augmentation strategies and three diverse encoder architectures, and b) employs an unsupervised linear classifier (K-means) in the detection stage, rendering it label-free. 

\section{Method}
\label{Method}

\subsection{Augmentations}
As previously emphasized,  the choice of augmentation methods is pivotal in shaping the \acrshort{ssl} objective, as it dictates what the non-contrastive \acrshort{ssl} models learn. In this section, we explain the considered augmentation methods.

\textbf{Swap Noise.}
Given a traffic network sample $i \sim D$ from dataset $D$ with $i \in \bbbr^{d_D}$, where $d_D$ is the number of features of sample $i$. To generate an augmented version of this sample $i'$, each feature of $i$ is randomly replaced with a feature at the same position from other samples in $D$ with probability $p$ sampled from a Bernoulli distribution \cite{ucar_subtab_2021,yoon_vime_2020,bahri_scarf_2022,somepalli_saint_2021}. This procedure is given by 
\begin{equation}
    i' = i \odot (1 - m) +  j \odot m ,
\end{equation}
where $m \in \{0, 1\}^{d_D}$ is a binary mask vector with each element drawn from a Bernoulli distribution, $j$ is a feature vector where each feature is randomly sampled from the original data within the same feature and $\odot$ is the element-wise multiplication. 

\textbf{Zero Out Noise.}
Similar to \textit{Swap Noise}, features are randomly replaced by zeros in this augmentation. The generation of an augmented sample $i'$ from a sample $i$ is then given by 
\begin{equation}
    i' = i \odot (1 - m) ,
\end{equation}
where $m$ is again a binary vector sampled from a Bernoulli distribution with parameter $p$ \cite{ucar_subtab_2021,lotfi_network_2023}.

\textbf{Gaussian Noise.}
In addition to these replacement methods, we can add Gaussian noise onto randomly selected feature values \cite{ucar_subtab_2021,mirza_self-supervision_2021}. The noise is sampled from a normal distribution with $\epsilon = N(\mu, \sigma^2)$, where $\mu \in \bbbr$ is the mean and $\sigma^2 \in \bbbr_{>0}$ is the variance. Formally, this is given by 
\begin{equation}
    i' = i + \overrightarrow{\epsilon} \odot m ,
\end{equation}
where $\overrightarrow{\epsilon}$ is a vector of values, with each value sampled from the normal distribution $N(\mu, \sigma^2)$ and $m$ is the binary mask vector sampled from a Bernoulli distribution with parameter $p$.

\textbf{Random Shuffle.}
The former augmentation methods alter the features' values. In contrast to the former augmentations, we can randomly shuffle the features' positions within a sample $i$ to generate the augmented version of the sample \cite{wang_network_2021}. For this shuffling, a version of the Fisher-Yates algorithm given in Appendix~\ref{app:fisher} is used to generate the augmented version of sample $i$. 

\textbf{Subsets.}
Instead of creating two views, in this augmentation the features of D are split into k subsets, before being fed into the encoder $f_\theta$ \cite{ucar_subtab_2021}. Each subset consists of a set of features that can overlap with a neighbor subset with a defined percentage of the subset's number of features. We randomly shuffle the dataset features at the beginning of each run to remove any negative bias created by the order of features before building the subsets. 

The subsets can be seen as different views of the original sample fed into the model. In contrast to the previously mentioned augmentation methods, \textit{Subsets} automatically creates more than one view and thus is not required to be executed multiple times. With $k > 2$ as the number of subsets, more than two views are obtained. Therefore, we compute the pairwise loss of all views and take the mean as the final loss. At test time, the samples must be split into subsets similar to those at training time, because the trained encoder is adjusted to the subset's shape. Therefore, each subset $s_1, s_2, \ldots, s_k$ is processed by the encoder $f_\theta$ to generate a representation for each subset $y_1, y_2, \ldots, y_k = f_\theta(s_1), f_\theta(s_2), \ldots, f_\theta(s_k)$. Then, these representations of the subsets are aggregated using the representation’s element-wise mean, forming the final representation for the downstream task.

\textbf{Mixup.}
Each former-explained augmentation method operates in the input space, implying that the augmentations take a raw network traffic sample and transform it into one or multiple views of this sample. In contrast, the \textit{Mixup} augmentation operates in the representation space \cite{yoon_vime_2020,somepalli_saint_2021}. Thus, the encoders simply receive two copies of the input sample and generate the representations $y = f_\theta(i), y' = f_{\theta}(i)$. Then, \textit{Mixup} creates a convex combination between $y$ and another randomly selected representation of the current batch $y_j$. This augmentation is mathematically described by

%Then, \textit{Mixup} is converging $y$ towards another randomly selected representation of the current batch $y_j$ which is mathematically described by 
\begin{equation}
    \bar{y} = \alpha * y + (1-\alpha)* y_j
\end{equation}
Similarly, the second representation $y'$ is augmented with a different randomly selected representation of the batch.

\begin{figure}
\includegraphics[width=\textwidth]{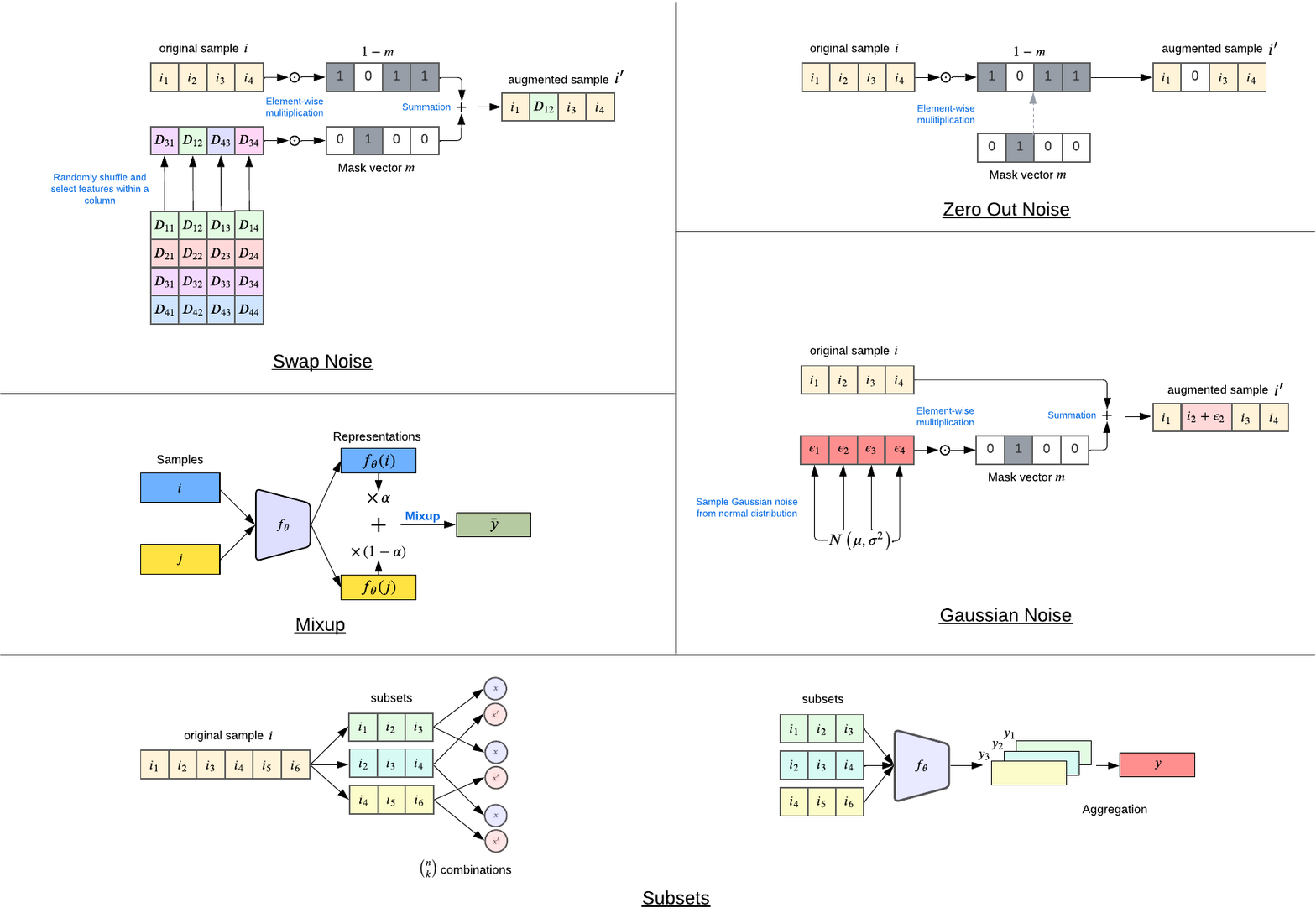}
%\includeg[width=\textwidth]{pics/Augmentations.pdf}
\caption{Visualisation of different augmentation strategies. Swap Noise: each feature of sample $i$ is randomly replaced with a feature from the same position in other samples, with probability $p$ from a Bernoulli distribution. $m$ is a binary mask vector with elements drawn from a Bernoulli distribution, and $\odot$ represents element-wise multiplication. Zero Out Noise: to generate the augmented view $i'$, features of sample $i$ are multiplied element-wise by 1 minus the binary mask vector. Gaussian Noise: $\overrightarrow{\epsilon}$ is a vector of values, with each value sampled from the normal distribution. This vector is element-wise multiplied with a binary mask vector and summed with the original sample vector $i$ to generate the augmented sample. Mixup: operates in the representation space where the encoder $f_\theta$ receives two copies of the representations $y = f_\theta(i)$ and $y' = f_{\theta}(i')$. Mixup creates a convex combination between $y$ and another randomly selected representation of the current batch $y_j$. Similarly, the second representation $y'$ is augmented with a different randomly selected representation of the batch. Subsets: dataset features are split into $k$ subsets before being fed into the encoder $f_\theta$. Each subset can overlap with a neighboring subset by a defined percentage of features. For $k > 2$, more than two views are obtained. Each subset is processed by the encoder $f_\theta$ to generate representations $y_1, y_2, \ldots, y_k$. These representations are aggregated using their element-wise mean, forming the final representation for the downstream task. } \label{fig:augmentations}
\end{figure}

The different augmentation strategies are further illustrated in Figure~\ref{fig:augmentations}.

\subsection{Encoders}
\begin{comment}
    Therefore, each subset $s_1, s_2, \ldots, s_k$ is processed by the encoder $f_\theta$ to generate a representation for each subset $y_1, y_2, \ldots, y_k = f_\theta(s_1), f_\theta(s_2), \ldots, f_\theta(s_k)$. Then, these representations of the subsets are aggregated using the representation’s element-wise mean, forming the final representation for the downstream task.
\end{comment}
\label{Encoders}

The representations generated by the \acrshort{ssl} model's encoder play a crucial role in improving the performance of the downstream task. To this end, three different encoder architectures are chosen to serve as the representation learning backbone for the non-contrastive \acrshort{ssl} models. The details of each architecture are explained in the following.

\textbf{CNN.}
For the \glspl{cnn} architecture, the procedure described in \cite{lotfi_network_2023} is adhered to, where the authors reshaped a network traffic sample with $d_D$ features to $H \times W \times C = 1 \times d_D \times 1$. Through this processing, we can use a $2D$~convolutional layer with filters of height $H = 1$ and arbitrary width $W$. The precise architecture is given in Appendix~\ref{tab:cnn}.

\textbf{MLP.}
The encoder consists of four fully connected layers, with the first layer being the input layer, followed by two hidden layers and the output layer. After each of the first three layers, we perform batch normalization followed by applying a ReLU activation function before feeding the output of the current layer to the next layer. The embedding dimension is set to $256$ for all layers. Consequently, the input dimension of the first input layer equals the number of features from the input sample, whereas all remaining layers have an input dimension of $256$. The MLP architecture is adapted from \cite{bahri_scarf_2022}. 

\textbf{FT-T.}
The \acrfull{ft} is a supervised transformer model introduced for tabular data consisting of a Feature Tokenizer and a Transformer \cite{gorishniy_revisiting_2023}. We decided to take advantage of the \acrshort{ft} because, unlike typical Transformers used e.g. in \cite{wang_network_2021}, the \acrshort{ft} is capable of handling numerical and categorical data. Given that the pre-training of the encoder is label-free, we exclude the classification token utilized by the prediction head of the \acrshort{ft} to integrate it into our pipeline. The Feature Tokenizer creates an embedding of an input network traffic sample by separately processing numerical and categorical features. Afterward, the embeddings of all features are stacked upon each other, which forms the final embedding of the input sample. The embedding is further processed by the Transformer part of the encoder, which consists of a stack of Transformer layers. The structural architecture of the Transformer layers remained unaltered, following the specifications in the original paper \cite{gorishniy_revisiting_2023}. In our implementation, we set the embedding dimension of the feature tokenizer to $32$ and the number of self-attention heads for the Multi-Head Self-Attention mechanism to $4$.
In addition, we chose $4$ Transformer layers, encoding the input embedding. The output of the Transformer encoder is then flattened for the subsequent computations. Furthermore, we added a dropout of $0.1$ into each attention and feedforward sub-layer, similar to the original implementation \cite{gorishniy_revisiting_2023}. 

The introduced encoders replace the original encoders in the implementations of the SSL models in our framework. This adaption allows the models to operate on network traffic data. To ensure a fair and competitive comparison, the architecture of the encoders described above is kept identical for each \acrshort{ssl} model.

\subsection{Non-Contrastive \acrshort{ssl} Models}
\label{sec:ssl-methods}

\begin{figure}
  \centering
 % \includesvg[width=\textwidth]{pics/ssl_overview}
 \includegraphics[width=\textwidth]{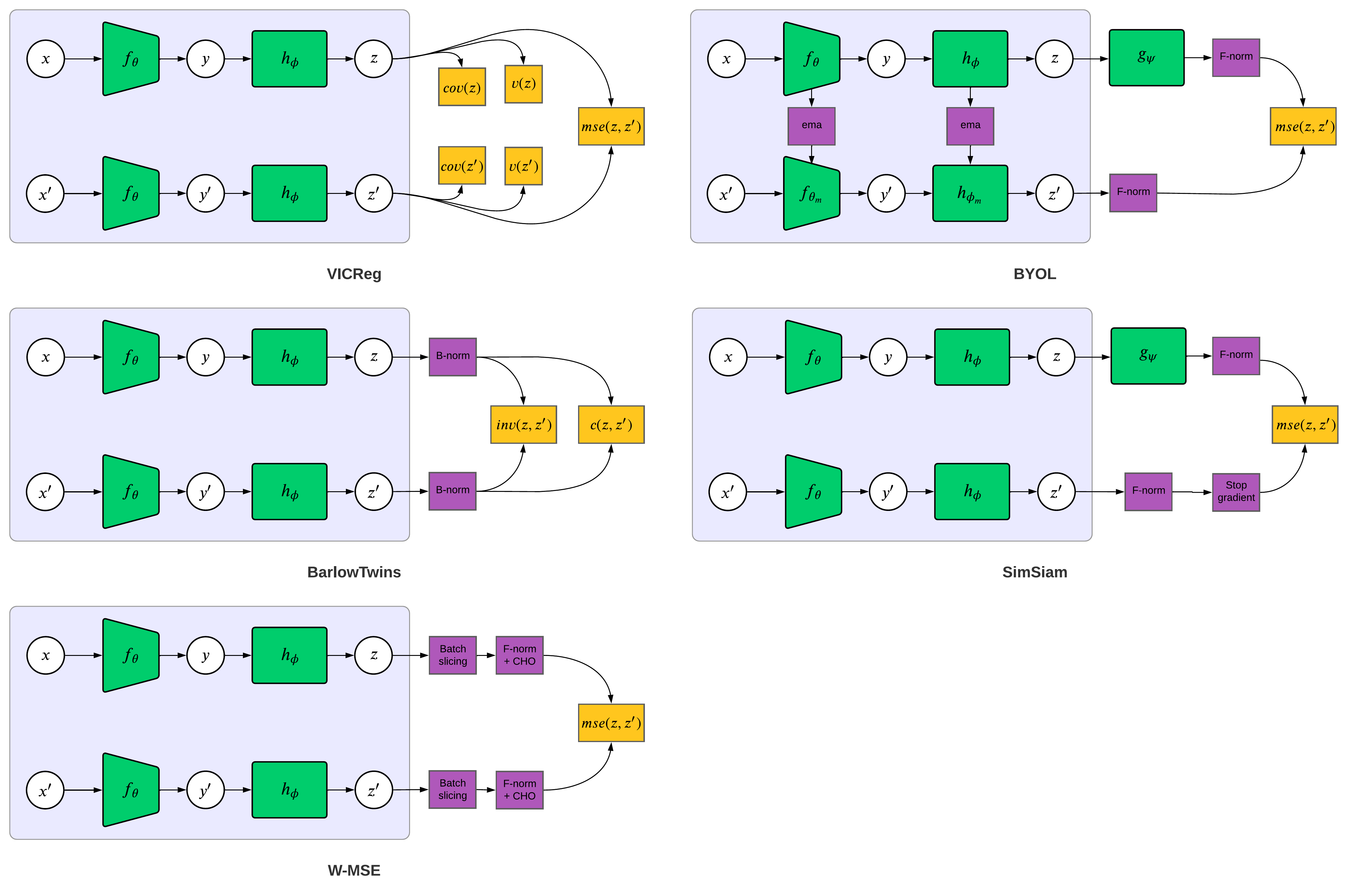}
  \caption{Comparison of the different non-contrastive \acrshort{ssl} models. The two augmented views $x$, $x'$ are fed to an encoder $f$ (can be an MLP, CNN, or FT-T ) with weights $\theta$ which yields the representations $y = f_{\theta}(x)$, $y' = f_{\theta}(x')$. Then, $y$ and $y'$ are further processed by the network $h$  with weights $\phi$. $h$ is an \acrshort{mlp} (two fully-connected layers with batch normalization and ReLU activation). After this step, different criteria are applied to the projector embeddings $z$ and $z'$. VICReg: regularizes the variance and covariance of each branch independently with $v$ and $cov$, respectively. The invariance term is determined as the mean-squared distance between each pair of vectors $z$ and $z'$. The final loss is the weighted sum of these three terms. BYOL: one branch incorporates an additional predictor, denoted as $g$ with weights $\psi$, to map the output of one network to the other, resulting in an asymmetric architecture. The output embeddings of the two branches are \acrfull{f-norm} and the similarity loss is computed as the \acrfull{mse} between them. Barlow Twins: its objective function assesses the cross-correlation matrix between the outputs of the two branches and has two terms: an invariance term ($inv$) that aims to set the diagonal elements of the cross-correlation matrix to 1 and a decorrelation term ($c$), which decorrelates pairs of different dimensions within the batch-wise normalized (B-Norm) embeddings. SimSiam: adds a predictor network in one branch and a stop-gradient operation in the other, omitting BYOL's moving average. W-MSE: applies batch slicing and a Cholesky decomposition-based whitening transformation to \acrshort{f-norm} embeddings. The loss is the \acrfull{mse} between whitened, normalized embeddings of the two branches.}\label{fig:ssl_overview}
\end{figure}

As mentioned in Section \ref{Introduction}, non-contrastive \acrshort{ssl} models minimize the distance of representations obtained from different augmented versions of a sample, which is realized through a joint embedding architecture. However, a shortcoming of joint embedding architectures is a phenomenon known as collapse, where the two branches ignore the inputs and produce constant output vectors (trivial solutions). In the this subsection, we delve into the distinct features of these models with a particular emphasis on their role in preventing collapse, which are illustrated in Figure \ref{fig:ssl_overview}.

 The two augmented views $x$, $x'$ are fed to  an encoder $f$ with weights $\theta$ which yields the representations $y = f_{\theta}(x)$, $y' = f_{\theta}(x')$. Then, $y$ and $y'$ are further processed by the network $h$ (referred to as the projector) with weights $\phi$. In our case, $f$  can be any one of three encoders described in Section \ref{Encoders} and $h$ is an \acrshort{mlp} (two fully-connected layers with batch normalization and ReLU activation) with embedding dimension 256 for each model to ensure a fair comparison. After this step, different criteria are applied to the projector embeddings $z$ and $z'$: 
 
 \textbf{BYOL.} Collapse is prevented through architectural modifications, where in one branch, the weights $\theta_{m}$ for the encoder $f$ and $\phi_{m}$ for the projector $h$ are the estimated moving averages of their respective weights $\theta$ and $\phi$ in the other branch. One branch incorporates an additional predictor, denoted as $g$ with weights $\psi$, to map the output of one network to the other, resulting in an asymmetric architecture. Finally, the output embeddings of the two branches are \acrfull{f-norm} \footnote{In our implementation, F-norm  always refers to $\ell_{2}$ normalization.}, and the similarity loss is computed as the \acrfull{mse} between them \cite{grill_bootstrap_2020}.
 
 \textbf{SimSiam.} The estimated moving average operation from BYOL is omitted because it was found to be unnecessary for preventing representation collapse \cite{chen_exploring_2020}. Similar to BYOL, SimSiam includes a predictor network in one branch and a stop-gradient operation in the other branch \cite{bardes2021vicreg}. The stop-gradient operation allows the branch with the predictor to be optimized with the projector output of the other branch as the target, but not the other way around. 
 
 \textbf{Barlow Twins.} The design of the objective function, rather than architectural modifications, is responsible for preventing collapse. The objective function assesses the cross-correlation matrix between the outputs of the two branches with the goal of minimizing the deviation from the identity matrix \cite{zbontar2021barlow}. It comprises two key terms: an invariance term ($inv$) that aims to set the diagonal elements of the cross-correlation matrix to 1 and a decorrelation term ($c$), which decorrelates pairs of different dimensions within the batch-wise normalized (B-Norm) embeddings \cite{bardes2021vicreg}, i.e., it aims to set the off-diagonal elements of the cross-correlation matrix to 0.   
 
 \textbf{VICReg.} Similar to Barlow Twins, VICReg avoids collapse through its objective function, which balances three essential components: a variance term, a covariance term, and an invariance term. The variance and covariance of each branch undergo independent regularization through $v$ and $cov$, respectively. The variance regularization term $v$ imposes a constraint on the variance along the batch dimension, ensuring it exceeds a specified threshold for every embedding dimension. Simultaneously, the covariance regularization term $cov$ is defined as the sum of the squared off-diagonal coefficients of the covariance matrix. Moreover, both the variance and covariance regularization terms are computed independently for each branch. The invariance term is determined as the mean-squared distance between each pair of vectors $z$ and $z'$. The final loss is the weighted sum of these three terms \cite{bardes2021vicreg}. 
 
 \textbf{W-MSE.} Similar to VICReg and Barlow Twins, W-MSE prevents collapse through its objective function. This involves a batch slicing operation that reorganizes batches of projector output embeddings $z$ and $z'$ into smaller sub-batches \cite{huang2018decorrelated}. Following this, a Cholesky decomposition-based whitening transformation is applied to the \acrshort{f-norm} embeddings of each sub-batch \cite{bardes2021vicreg}. Finally, the loss is computed as the mean squared error between the whitened, normalized embeddings of the two branches \cite{ermolov2021whitening}.

\subsection{Classifier}
\begin{figure}
%\includesvg[width=\textwidth]{pics/Classifier}
\includegraphics[width=\textwidth]{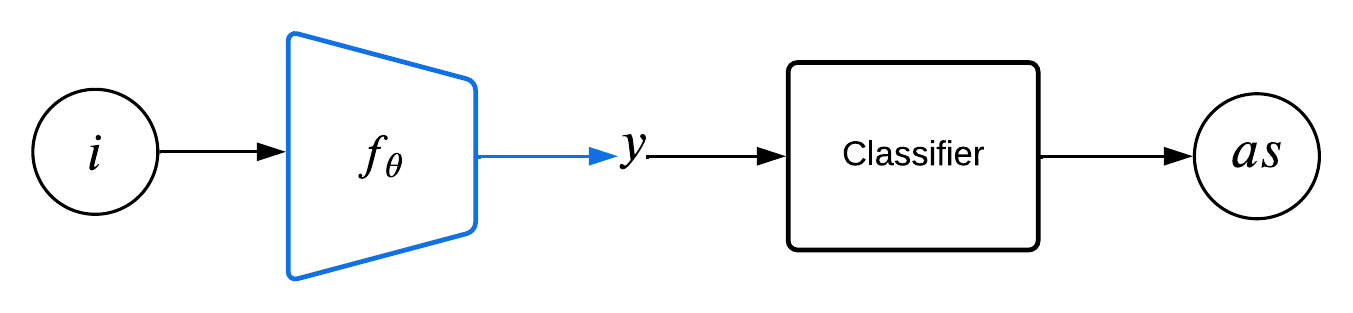}
\caption{K-means classifier generates an anomaly score ($as$) for a network traffic sample $i$.} \label{fig:classifier}
\end{figure}

After the training phase of the non-contrastive \acrshort{ssl} models has completed, only the encoder $f_\theta$ is kept, and all other parts of each model are discarded \cite{jaiswal_survey_2021}. The weights of the encoder are frozen, and a simple classifier is trained on top of the frozen representation of the data, as shown in Figure \ref{fig:classifier}. Training a linear classifier on top of a pre-trained encoder and using a labeled set of the dataset for fine-tuning on the downstream task is the most common approach in computer vision, which is also used in \acrshort{nid} by \cite{wang_network_2021,yang2022android,towhid2022encrypted,menon2023vicra}. However, in this paper, no access to labeled data for fine-tuning is assumed, and the learned representations of the encoders are evaluated via a simple K-means algorithm with a single cluster center in an unsupervised manner, following \cite{sehwag_ssd_2021}. The K-means classifier trained on top of the encoder calculates an anomaly score from a given feature representation $y = f_\theta(i)$. The anomaly score $ac$ is defined as the Euclidean distance $E$ between the cluster center and the representation of the test sample $ts$ processed by the encoder $f_\theta$ given as
\begin{equation}
    as = \left\| E \right\|_2 = \left\| f_\theta(ts) - cc \right\|_2
\end{equation}
where cluster center $cc$ is given by: 
\begin{equation}
    cc = \frac{1}{n_D} \sum_{j=1}^{n_D} f_\theta(i_j) 
\end{equation}
$D = \{i_1, i_2, \ldots, i_{n_D}\}$ are the $n_D$ data samples used for training the unsupervised classifier.

\section{Experiments}
\label{Experiment}
In the preceding sections, the four key components of our pipeline were explained: the augmentation methods, encoder architectures, non-contrastive \acrshort{ssl} models, and the unsupervised classification head, yielding a complete pipeline suitable for a \acrshort{nid} task. We treat every unique combination as an individual experiment, resulting in a total of $90$ experiments ($3 \times 6 \times 5$). Each combination is then hyper-optimized, trained, and evaluated on two \acrshort{nids} datasets under identical conditions to ensure a fair performance comparison. In this section, the evaluation protocol, metrics, datasets, and hyperparameter optimization strategy are examined. The complete pipeline is given in our repository\footnote{\url{https://github.com/renje4z335jh4/non_contrastive_SSL_NIDS}}.  

\subsection{Evaluation Protocol and Metrics}
Ensuring a fair and consistent comparison of different models is crucial. Unfortunately, in scientific publications, this is not always achieved due to inconsistent hyperparameter tuning or the use of misleading metrics for evaluation, such as accuracy in an unbalanced test set. Moreover, various choices for the class of interest, discrepancies in the training protocol, and different ratios of anomalies in the training set are barriers to comparative evaluations. Therefore, we adhere to the evaluation protocol of \cite{alvarez_revealing_2022}, where the training set consists exclusively of normal data and the test set includes both normal data and anomalies, with the split ratio as specified in \cite{alvarez_revealing_2022}. The positive class is consistently defined as the anomalous class, which forms the basis for the performance metrics. These include the threshold-independent metric AUROC and threshold-dependent metrics, precision, recall (detection rate), and F1-score, computed with an optimal threshold. To account for statistical uncertainty, each combination of augmentation method, encoder architecture, and \acrshort{ssl} model is treated as a distinct experiment and executed in 10 runs. The mean and standard deviation are subsequently calculated over all runs, ensuring robust performance evaluations and more accurate estimations of each model's performance. 

\subsection{Datasets}
\begin{table}[!h]
    \caption[datasets]{General information on the datasets after preprocessing,}
    \label{tab:datasets}
    \centering
    \begin{tabular}{|l|S[table-format=8,table-number-alignment=left]|S[table-format=4,table-number-alignment=left]|S[table-format=2.4,table-number-alignment=left]|}
        \hline 
         \textbf{Dataset} & \textbf{Number of Samples} & \textbf{Number of Features} & \textbf{Attack Ratio}    \\ \hline
         \acrshort{unsw} & 154098 & 196 & 0.4437 \\
         \acrshort{5g} & 1215655 & 58 & 0.6072 \\
         \hline
    \end{tabular}
\end{table}
To benchmark the different \acrshort{ssl} models in the pipeline, two \acrshort{nids} datasets are utilized:
\begin{comment}
For benchmarking the different \acrshort{ssl} models of our pipeline, we used two \acrshort{nids} datasets 
\end{comment}
 \acrshort{unsw} \cite{moustafa_unsw-nb15_2015} and \acrshort{5g} \cite{samarakoon_5g-nidd_2022}. The \acrshort{unsw} dataset is well-recognized in the \acrshort{nid} domain, proving more suitable for modern \acrshort{nid} than the NSL-KDD dataset  \cite{divekar_benchmarking_2018}. The \acrshort{5g} dataset is collected using the 5G Test Network (5GTN) in Finland. A key feature of this dataset is the generation of benign traffic by actual mobile devices in the network, as opposed to simulated traffic. The benign traffic consists of HTTP, HTTPS, SSH, and SFTP traffic. In addition, it includes two attack categories: Port Scan (including SYN Scan, TCP Connect Scan, UDP Scan) and Dos/DDoS (covering ICMP flood, UDP flood, SYN flood, HTTP flood, Slow rate DoS - Slowloris, Slow rate DoS - Torshammer) attacks.

Both datasets were initially cleaned by removing NaN values, dropping duplicated features and samples, and normalizing the values of the features. Furthermore, categorical features were one-hot encoded. Thus, the datasets only consist of numerical features that fit each encoder's input type. Finally, the malicious samples were merged for the \acrshort{nid} task. A script for this general preprocessing is provided in our repository\footnote{\url{https://github.com/renje4z335jh4/non_contrastive_SSL_NIDS/blob/main/src/data/process.py}} \footnote{Aside from the feature preprocessing steps described above, no further feature selection methods were applied}. Table~\ref{tab:datasets} summarizes the information of the datasets after the preprocessing step.

\subsection{Hyperparameter Optimization}
\label{Hyperparameter Optimization}

{
\sisetup{output-exponent-marker=\ensuremath{\mathrm{e}}}
The implementation comprises three sets of variable hyperparameters: model-specific parameters, augmentation parameters, and general training parameters such as learning rate and epochs. Due to the absence of established references guiding the appropriate configuration of these parameters within our pipeline, and recognizing that directly adopting hyperparameter values from original papers may result in biased and non-competitive outcomes, optimization of hyperparameters was conducted for each dataset, model, augmentation, and encoder combination using Tune \cite{liaw2018tune}.

The ADAM optimizer was consistently employed for model optimization both in hyperparameter optimization and final runs. During the initial optimization phase, the learning rate was set to $\num{1e-4}$ and the maximum number of epochs was fixed at $200$. Subsequently, optimal model and augmentation parameters were determined using BayesOptSearch \cite{bayesian_nogueria} with $200$ trials, except for the W-MSE model and subset augmentation, which utilized a BasicVariantGenerator due to an integer search space. Following this, the identified optimal parameters were established, and a grid search was conducted on common learning rates ($\num{1e-2}$, $\num{1e-3}$, $\num{1e-4}$, $\num{1e-5}$), each with three trials. Finally, the optimal number of epochs was determined by training the models with the previously obtained parameters for $200$ epochs and three runs. The epoch with the highest average metric over the three runs was considered as the optimal number of epochs. Optimal hyperparameters for each combination are available in our repository\footnote{\url{https://github.com/renje4z335jh4/non_contrastive_SSL_NIDS/blob/main/hyperopt/best_config.yml}}.

}

\subsection{Experimental Environment}

A $64$-Bit computer with Debian version $11.7$ is used to execute the experiments. As hardware components, the Intel(R) Core(TM) i$9$-$10980$XE CPU \@$3.00$GHz with $32$~GB RAM and an NVIDIA RTX A$5000$ GPU with $24$~GB VRAM are given. %All models are implemented in PyTorch \cite{paszke_pytorch_2019} $1.13.1$.

\section{Results}
\label{Results}
Our experimental results, detailed in Tables \ref{tab:comparison_unswnb15} and \ref{tab:comparison_5gnidd}, showcase the combinations of augmentation strategy and encoder architecture that yielded the highest average precision, recall, F1-score, and AUCROC metrics for each \acrshort{ssl} model on the \acrshort{unsw} and \acrshort{5g} datasets, respectively. 
   
\textbf{\acrshort{unsw}.} For the \acrshort{unsw} dataset, BYOL exhibits comparably lower performance than other models. Notably, the \textit{Gaussian Noise} augmentation strategy, which was previously deemed unsuitable by \cite{wang_network_2021} yields the best result for BYOL! On the contrary, the \textit{Random Shuffle} augmentation, proposed in \cite{wang_network_2021}, consistently underperforms when combined with any encoder or \acrshort{ssl} model. As a result, it is not featured in Table \ref{tab:comparison_unswnb15}. The poor performance of the \textit{Random Shuffle} augmentation, compared to other augmentation methods, becomes more evident in Tables \ref{tab:comparison_BYOL_UNSW} and \ref{tab:comparison_SimSiam_UNSW}, where the performance of BYOL and SimSiam models are compared across all augmentation methods on the \acrshort{unsw} dataset. Another absent augmentation strategy in Table \ref{tab:comparison_unswnb15} is \textit{Swap Noise}, used by the authors in \cite{menon2023vicra} in conjunction with VICReg and an \acrshort{mlp} encoder. In our experiments, VICReg attains the highest average precision, F1-Score, and AUCROC with the \textit{Subsets} augmentation. Similarly, Barlow Twins demonstrates the best average performance metrics when combined with the \acrshort{mlp} encoder and the \textit{Subsets} augmentation method. In addition, the \textit{Zero Out Noise} augmentation, combined with the SimSiam model and \acrshort{ft}, yields the highest detection rate. 

\begin{table}[h]
\caption{Comparison of non-contrastive SSL models with the highest average performance metrics (all with standard deviation) on the \acrshort{unsw} dataset.}
\label{tab:comparison_unswnb15}
\centering
\sisetup{separate-uncertainty=true,text-series-to-math = true}
\begin{tabular}{|lll|S[table-format=2.3(4),table-number-alignment=left]|S[table-format=2.3(4),table-number-alignment=left]|S[table-format=2.3(4),table-number-alignment=left]|S[table-format=2.3(4),table-number-alignment=left]|}
\hline
\textbf{Model} & \textbf{ENC} & \textbf{AUG} & \textbf{Precision} & \textbf{Recall} & \textbf{F1-Score} & \textbf{AUROC}  \\ \hline
BYOL & FT-T & GN &  0.720\pm0.021  &  0.776\pm0.024  &  0.747\pm0.022 &  0.704\pm0.037 \\
SimSiam & FT-T & ZON &  0.762\pm0.049  &  {\textbf{\num{0.823\pm0.053}}}  &  0.791\pm0.051 &  0.762\pm0.048 \\
VICReg & MLP & S &  {\textbf{\num{0.788\pm0.059}}}  &  0.810\pm0.062  &  {\textbf{\num{0.798\pm0.056}}} &  {\textbf{\num{0.786\pm0.071}}} \\
BarlowTwins & MLP & S &  0.783\pm0.066  &  0.809\pm0.053  &  0.795\pm0.056 &  0.764\pm0.105 \\
W-MSE & CNN & M &  0.763\pm0.031  &  0.806\pm0.019  &  0.784\pm0.018 &  0.756\pm0.043 \\
\hline
\end{tabular}
\end{table}

\begin{table}[h]
\caption{Comparison of non-contrastive SSL models with the highest average performance metrics (all with standard deviation) on the \acrshort{5g} dataset.}
\label{tab:comparison_5gnidd}
\centering
\sisetup{separate-uncertainty=true,text-series-to-math = true}
\begin{tabular}{|lll|S[table-format=2.3(4),table-number-alignment=left]|S[table-format=2.3(4),table-number-alignment=left]|S[table-format=2.3(4),table-number-alignment=left]|S[table-format=2.3(4),table-number-alignment=left]|}
\hline
\textbf{Model} & \textbf{ENC} & \textbf{AUG} & \textbf{Precision} & \textbf{Recall} & \textbf{F1-Score} & \textbf{AUROC}  \\ \hline
BYOL & CNN & SN &  0.867\pm0.015  &  0.911\pm0.017  &  0.888\pm0.013 &  0.775\pm0.017 \\
SimSiam & CNN & S &  0.841\pm0.070  &  0.877\pm0.070  &  0.858\pm0.069 &  0.724\pm0.134 \\
VICReg & CNN & GN &  {\textbf{\num{0.932\pm0.010}}}  &  {\textbf{\num{0.961\pm0.026}}}  &  {\textbf{\num{0.946\pm0.009}}} &  0.908\pm0.005 \\
BarlowTwins & MLP & M &  0.916\pm0.012  &  0.909\pm0.012  &  0.912\pm0.009 &  {\textbf{\num{0.925\pm0.009}}} \\
W-MSE & MLP & M &  0.836\pm0.054  &  0.891\pm0.057  &  0.863\pm0.056 &  0.756\pm0.077 \\
\hline
\end{tabular}
\end{table}

\textbf{\acrshort{5g}.} VICReg, combined with a \acrshort{cnn} encoder and the \textit{Gaussian Noise} augmentation, achieved the highest average precision, recall, and F1-Score among all other non-contrastive \acrshort{ssl} models. This showcases the viability of \textit{Gaussian Noise} as an augmentation strategy. One possible reason for the exclusion of this strategy in \cite{wang_network_2021} could be attributed to either insufficient hyperparameter optimization or its combination with other \acrshort{cv} augmentations, such as horizontal flip, vertical flip, random crop, or even \textit{Random Shuffle}, leading to suboptimal performance. Similar to the results in Table \ref{tab:comparison_unswnb15}, \textit{Random Shuffle} did not achieve competitive results, irrespective of the \acrshort{ssl} model it was employed with. Consequently, it is not featured in Table \ref{tab:comparison_5gnidd} as well. Barlow Twins, in conjunction with the \acrshort{mlp} encoder and \textit{Mixup} augmentation, achieved the highest AUCROC. In addition, \textit{Mixup} is the augmentation method that, along with the \acrshort{mlp} encoder, achieved the highest metrics for W-MSE on this dataset.

Notably, \textit{Mixup} is the only augmentation method used to operate in the representation space, resulting in competitive outcomes on both datasets. This suggests that incorporating augmentation in the representation space can serve as a practical alternative to augmentations in the input space.
Furthermore, the \acrshort{ft} encoder, which, in combination with BYOL and the SimSiam model, achieved the highest average performance metrics for these models in Table \ref{tab:comparison_unswnb15}, is notably absent in Table \ref{tab:comparison_5gnidd}. i.e., for all other combinations of augmentation and \acrshort{ssl} models, the \acrshort{cnn} and \acrshort{mlp} encoders consistently outperform the \acrshort{ft} on both datasets. This observation indicates that the choice and optimization of augmentation techniques and hyperparameters may exert a more significant influence than the use of deeper and more complex architectures as backbone encoders. Experimental findings presented in \cite{yang2022android} align with this perspective, indicating that employing deeper ResNet architectures as encoders in their BYOL model for android malware detection resulted in diminishing returns in accuracy.

Moreover, comparing the results in Tables \ref{tab:comparison_unswnb15} and \ref{tab:comparison_5gnidd} shows that when the augmentation involves \textit{Subsets}, irrespective of the \acrshort{ssl} model or encoder used, it leads to high uncertainty across performance metrics. Another drawback of this augmentation is its computational complexity during training time, as the computation of loss involves combinations of projections, which limits the number of subsets for data splitting \cite{ucar_subtab_2021}.

As detailed in Section \ref{sec:ssl-methods}, BYOL and SimSiam share a similar architecture, differing primarily in the absence of an estimated moving average in SimSiam compared to BYOL. This architectural distinction is a key feature that sets the two models apart. To emphasize the differences between these two \acrshort{ssl} models, we also present the performance metrics for both models with the FT-Transformer as its encoder and integrate all augmentation strategies on the \acrshort{unsw} dataset. The corresponding results are detailed in Table  \ref{tab:comparison_BYOL_UNSW} and \ref{tab:comparison_SimSiam_UNSW}. 
\begin{table}[h]
\caption{Results of the BYOL model with the FT-T as the encoder utilizing different augmentation strategies on the \acrshort{unsw} dataset.}
\label{tab:comparison_BYOL_UNSW}
\centering
\sisetup{separate-uncertainty=true,text-series-to-math = true}
\begin{tabular}{|lll|S[table-format=2.3(4),table-number-alignment=left]|S[table-format=2.3(4),table-number-alignment=left]|S[table-format=2.3(4),table-number-alignment=left]|S[table-format=2.3(4),table-number-alignment=left]|}
\hline
\textbf{Model} & \textbf{ENC} & \textbf{AUG} & \textbf{Precision} & \textbf{Recall} & \textbf{F1-Score} & \textbf{AUROC}  \\ \hline
BYOL & FT-T & M &  0.679\pm0.014  &  0.733\pm0.014  &  0.705\pm0.014 &  0.644\pm0.039 \\
BYOL & FT-T & RS &  0.625\pm0.005  &  0.676\pm0.006  &  0.650\pm0.005 &  0.520\pm0.012 \\
BYOL & FT-T & S &  0.679\pm.023  &  0.733\pm0.026  &  0.705\pm0.024 &  0.634\pm0.043 \\
BYOL & FT-T & GN &  {\textbf{\num{0.720\pm0.020}}}  &  {\textbf{\num{0.775\pm0.023}}}  &  {\textbf{\num{0.746\pm0.022}}} &  {\textbf{\num{0.703\pm0.037}}} \\
BYOL & FT-T & SN &  0.671\pm0.007  &  0.725\pm0.010  &  0.697\pm0.008 &  0.629\pm0.013 \\
BYOL & FT-T & ZON &  0.680\pm0.016  &  0.738\pm0.016  &  0.708\pm0.016 &  0.649\pm0.029 \\
\hline
\end{tabular}
\end{table}
%%%%%%%%%%%%%%%%%%%%%%%%%%%%%%%%%%%%%%%%%%%%%%%%%%%%%%%%

%%%%%%%%%%%%%%%%%%%%%%%%%%%%%%%%%%%%%%%%%%%%%%%%%%%%%%%%
%%%%%%%%%%%%%%%%%%%%%%%%%%%%%%%%%%%%%%%%%%%%%%%%%%%%%%%%

\begin{table}[ht]
\caption{Results of the SimSiam model with the FT-T as the encoder utilizing different augmentation strategies on the \acrshort{unsw} dataset.}
\label{tab:comparison_SimSiam_UNSW}
\centering
\sisetup{separate-uncertainty=true,text-series-to-math = true}
\begin{tabular}{|lll|S[table-format=2.3(4),table-number-alignment=left]|S[table-format=2.3(4),table-number-alignment=left]|S[table-format=2.3(4),table-number-alignment=left]|S[table-format=2.3(4),table-number-alignment=left]|}
\hline
\textbf{Model} & \textbf{ENC} & \textbf{AUG} & \textbf{Precision} & \textbf{Recall} & \textbf{F1-Score} & \textbf{AUROC}  \\ \hline
SimSiam & FT-T & M &  0.657\pm0.023  &  0.711\pm0.024  &  0.68\pm0.024 &  0.591\pm0.042 \\
SimSiam & FT-T & RS &  0.657\pm0.044  &  0.711\pm0.046  &  0.683\pm0.045 &  0.551\pm0.047 \\
SimSiam & FT-T & S &  0.664\pm0.058  &  0.718\pm0.062  &  0.69\pm0.059 &  0.569\pm0.630 \\
SimSiam & FT-T & GN &  0.720\pm0.021  &  0.777\pm0.020 &  0.747\pm0.020 &  0.712\pm0.039 \\
SimSiam & FT-T & SN &  0.672\pm0.075  &  0.714\pm0.062  &  0.692\pm0.066 &  0.597\pm0.093 \\
SimSiam & FT-T & ZON &  {\textbf{\num{0.761\pm0.048}}}  &  {\textbf{\num{0.823\pm0.053}}}  &  {\textbf{\num{0.791\pm0.050}}} &  {\textbf{\num{0.762\pm0.048}}} \\
\hline
\end{tabular}
\end{table}

In Table \ref{tab:comparison_BYOL_UNSW}, notable variations in AUCROC are observed for the BYOL model depending on the augmentation method. Specifically, employing \textit{Random Shuffle} leads to an AUCROC of $0.52$, while utilizing \textit{Gaussian Noise} significantly improves the AUCROC to $0.70$. A similar trend is evident in the performance of SimSiam, as presented in Table \ref{tab:comparison_SimSiam_UNSW}. When \textit{Random Shuffle} is employed, an AUCROC of $0.55$ is attained. However, opting for the \textit{Gaussian Noise} augmentation with SimSiam results in higher average performance metrics compared to the BYOL model utilizing the same augmentation method. 

\textbf{Comparison with unsupervised baselines:} Before concluding with the results section, the aim is to compare the non-contrastive models with the highest average performance metrics against two well-known unsupervised methods: DeepSVDD and AE. DeepSVDD, a deep learning-based one-classification method, entails mapping input data into a hypersphere with the objective of minimizing the hypersphere's volume. This mapping situates normal samples inside the hypersphere, while anomalies reside outside. AE, a vanilla autoencoder with a reconstruction-based objective, employs an \acrshort{mlp} architecture for the encoder and decoder. It is noteworthy that AE has demonstrated superior performance over other sophisticated reconstruction-based methods across diverse datasets, including \acrshort{nid} datasets, as detailed in \cite{alvarez_revealing_2022}. 
To ensure fairness and comparability of results, the hyperparameters of the baselines were tuned using Tune, following the same procedure described in Section ~\ref{Hyperparameter Optimization}. The results of this comparison are outlined in Tables \ref{tab:comparison_unswnb15_unsupervised} and \ref{tab:comparison_5g_unsupervised} for the \acrshort{unsw} and \acrshort{5g} datasets, respectively. In the \acrshort{unsw} dataset, VICReg achieves the highest average precision, while for other metrics, the AE consistently outperforms non-contrastive \acrshort{ssl} models. DeepSVDD shows less favorable results. A similar pattern is observed for the \acrshort{5g} dataset, where AE attains the highest average performance metrics. Although the highest average performance metrics achieved by the non-contrastive \acrshort{ssl} models are comparable to those of AE, the results underscore the significance of tuning baseline hyperparameters. Previous studies investigating the performance of non-contrastive \acrshort{ssl} models often fail to specify the extent of hyperparameter tuning in their baseline comparisons, potentially creating a misleading sense of confidence. 

\begin{table}[ht]
\caption{Comparison of non-contrastive \acrshort{ssl} models with the highest average performance metrics  (all with standard deviation) on the \acrshort{unsw} dataset against unsupervised models on the same dataset.}
\label{tab:comparison_unswnb15_unsupervised}
\centering
\sisetup{separate-uncertainty=true,text-series-to-math = true}
\begin{tabular}{|lll|S[table-format=2.3(4),table-number-alignment=left]|S[table-format=2.3(4),table-number-alignment=left]|S[table-format=2.3(4),table-number-alignment=left]|S[table-format=2.3(4),table-number-alignment=left]|}
\hline
\textbf{Model} & \textbf{ENC} & \textbf{AUG} & \textbf{Precision} & \textbf{Recall} & \textbf{F1-Score} & \textbf{AUROC}  \\ \hline
SimSiam & FT-T & ZON &  0.762\pm0.049  &  0.823\pm0.053  &  0.791\pm0.051 &  0.762\pm0.048 \\
VICReg & MLP & S &  {\textbf{\num{0.788\pm0.059}}}  &  0.810\pm0.062  & 0.798\pm0.056 &  0.786\pm0.071 \\
DeepSVDD & \textemdash & \textemdash &  0.683\pm0.021  &  0.735\pm0.025  &  0.708\pm0.023 &  0.656\pm0.047 \\
AE & \textemdash & \textemdash &  0.786\pm0.013  &  {\textbf{\num{0.837\pm0.029}}}  &  {\textbf{\num{0.811\pm0.018}}} &  {\textbf{\num{0.793\pm0.024}}} \\
\hline
\end{tabular}
\end{table}

\begin{table}[ht]
\caption{Comparison of non-contrastive \acrshort{ssl} models with the highest average performance metrics  (all with standard deviation) on the \acrshort{5g} dataset against unsupervised models on the same dataset.}
\label{tab:comparison_5g_unsupervised}
\centering
\sisetup{separate-uncertainty=true,text-series-to-math = true}
\begin{tabular}{|lll|S[table-format=2.3(4),table-number-alignment=left]|S[table-format=2.3(4),table-number-alignment=left]|S[table-format=2.3(4),table-number-alignment=left]|S[table-format=2.3(4),table-number-alignment=left]|}
\hline
\textbf{Model} & \textbf{ENC} & \textbf{AUG} & \textbf{Precision} & \textbf{Recall} & \textbf{F1-Score} & \textbf{AUROC}  \\ \hline
VICReg & CNN & GN &  0.932\pm0.010  &  0.961\pm0.026 &  0.946\pm0.009 &  0.908\pm0.005 \\
BarlowTwins & MLP & M &  0.916\pm0.012  &  0.909\pm0.012  &  0.912\pm0.009 &  0.925\pm0.009 \\
DeepSVDD & \textemdash & \textemdash &  0.895\pm0.060  &  0.937\pm0.055  &  0.915\pm0.057 &  0.865\pm0.117 \\
AE & \textemdash & \textemdash &  {\textbf{\num{0.939\pm0.027}}}  &  {\textbf{\num{0.965\pm0.018}}}  &  {\textbf{\num{0.951\pm0.020}}} &  {\textbf{\num{0.932\pm0.020}}} \\
\hline
\end{tabular}
\end{table}

%%%%%%%%%%%%%%%%%%%%%%%%%%%%%%%%%%%%%%%%%%%%%%%%%%%%%%%%%%
\section{Conclusion}
\label{Conclusion}

In this paper, we explore the interplay between augmentation methods, encoders, and non-contrastive \acrshort{ssl} models. We propose a two-stage pipeline: first, learning a useful representation of normal network traffic in a self-supervised manner; second, freezing the pre-trained encoder weights and using a K-means algorithm to distinguish between benign and attack data.
Our empirical findings revealed the poor performance of the \textit{Random Shuffle} method across all \acrshort{ssl} models. In contrast, \textit{Gaussian Noise}, previously deemed unsuitable by \cite{wang_network_2021}, yielded the best average performance metric for the BYOL model on the \acrshort{unsw} dataset. This could be due to insufficient hyperparameter optimization or its combination with other \acrshort{cv} augmentations, such as horizontal flip, vertical flip, random crop, or even \textit{Random Shuffle}, leading to suboptimal performance. \textit{Mixup} combined with Barlow Twins achieved the highest AUCROC on the \acrshort{5g} dataset, highlighting the effectiveness of augmentation in the representation space as an alternative to augmentations in the sample space. \textit{Subsets}, used with three different \acrshort{ssl} models on both datasets, exhibited competitive performance. Notably, in combination with VICReg, it achieved the highest precision, F1-score, and AUCROC on the \acrshort{unsw} dataset, though this method led to increased uncertainty in performance metrics. \textit{Zero Out Noise} and \textit{Swap Noise} showed competitive performance only with BYOL and SimSiam models, respectively, and only on the \acrshort{unsw} dataset. 

While our experiments underscore the importance of augmentation methods, they also reveal two significant drawbacks: a) these methods are not specifically tailored for \acrshort{nid}, and b) they do not necessarily satisfy the domain constraints of \acrshort{nid}, meaning they are not function-preserving and may lead to the generation of unrealistic samples \cite{sheatsley2021robustness}. Designing \acrshort{nid}-specific augmentation methods that satisfy domain constraints for \acrshort{ssl} methods is a promising avenue for future research.

Regarding SSL models, VICReg and Barlow Twins consistently attained higher average performance metrics than other \acrshort{ssl} models. The asymmetric architectural design choices in SimSiam and BYOL did not offer an advantage over these methods. For encoders, the \acrshort{ft} demonstrated competitive performance only as the backbone for BYOL and SimSiam models, and only on the \acrshort{unsw} dataset. In all other cases, the conceptually simpler \acrshort{mlp} and \acrshort{cnn} architectures proved more viable. 

Finally, this paper compares the performance of non-contrastive \acrshort{ssl} models to DeepSVDD and AE. The results show that non-contrastive \acrshort{ssl} models outperformed DeepSVDD, while AE achieved higher average performance metrics than non-contrastive \acrshort{ssl} models. This difference may be due to the use of a naive K-means detector. Future work could explore the utilization of improved distance metrics, such as the Mahalanobis distance \cite{mahalanobis1936generalized}, in the K-means detector. Additionally, incorporating more sophisticated unsupervised detectors, such as Isolation Forest or OCSVM, might address the minor performance gap observed between non-contrastive \acrshort{ssl} models and reconstruction-based approaches.

\appendix

\section{Encoder Structure}
\begin{table}
    \caption[Architecture of the \acrshort{cnn} encoder.]{Architecture of the \acrshort{cnn} encoder. \textit{conv} is a convolutional layer followed by a ReLU activation function, and pooling represents a pooling layer. For each layer, the kernel size, number of filters (only for convolutional layers), input shape, and output shape are given for an example network traffic sample with $196$ features.}
    \label{tab:cnn}
    \centering
    \begin{tabular}{|l|l|l|l|l|}
        \hline 
         \textbf{Layer} & \textbf{Kernel} & \textbf{Filter} & \textbf{Input}  & \textbf{Output}  \\ \hline
         conv1 & $1\times2$ & $32$ & $1 \times 196 \times 1 $ &  $1 \times 195 \times 32 $ \\
         conv2 & $1\times2$ & $64$ & $1 \times 195 \times 32 $ &  $1 \times 194 \times 64 $ \\
         conv3 & $1\times2$ & $128$ & $1 \times 194 \times 64 $ &  $1 \times 193 \times 128 $ \\
         pooling & $1\times3$ & $-$ & $1 \times 193 \times 128 $ &  $1 \times 64 \times 128 $ \\
         conv4 & $1\times2$ & $256$ & $1 \times 64 \times 128 $ &  $1 \times 63 \times 256 $ \\
         pooling & $1\times2$ & $-$ & $1 \times 63 \times 256 $ &  $1 \times 31 \times 256 $ \\
         conv5 & $1\times2$ & $512$ & $1 \times 31 \times 256 $ &  $1 \times 30 \times 512 $ \\
         pooling & $1\times4$ & $-$ & $1 \times 30 \times 512 $ &  $1 \times 7 \times 512 $ \\\hline
    \end{tabular}
    
\end{table}

\section{Augmentation}
\label{app:fisher}
\begin{algorithm}
    \caption{Pseudocode of Fisher-Yates inspired \textit{Random Shuffle} augmentation method.}
    \label{alg:fisheryates}
    \begin{algorithmic}
        \Require $i_j = \{f_j^{(1)}, f_j^{(2)}, \ldots, f_j^{(d_D)}\}$ \Comment{network traffic sample $i_j$ composed of $d_D$ features}
        \For{$k=d_D-1$ to $0$}
            \State $p \gets random\_integer(0, k)$
            \State $i_j^{(p)}, i_j^{(k)} = i_j^{(k)}, i_j^{(p)}$
        \EndFor
    \end{algorithmic}
\end{algorithm}

\bibliographystyle{splncs04}
\bibliography{Paper}

\end{document}